\documentclass[12pt]{article}
\usepackage{natbib}
\begin{document}

\author{D. Schmeltzer\\
Physics Department\\ City College of the City University of New York \\
Ph.D Program in Physics \\
New York, NY 10031}
\title{Hamiltonian Renormalization Groups}
\date{}
\maketitle

\begin{large}
I. Qualitative Introduction to Renormalization Groups (RG) \\
\end{large}

The RG has been introduced originally in order to handle
ultraviolet divergencies in Quantum Electrodynamics.  In the late
sixties it has been realized by Fisher and Kadanoff (1) that
critical exponents in phase transition can be computed within a RG
calculation with a finite cutoff $\Lambda$.  In the seventies the
method has been applied to solve the Kondo problem (2,3).  In the
eighties it has become clear that quantum phase transition at
$T=0$ and Quantum critical behavior can be studied within the RG
method. Today the RG method has become the major tool for
investigation of strongly correlated systems at low temperature.
In many problems we know the Hamiltonian but the Lagrangian might
be complicated. Since the original RG method is based on mapping
the quantum problem to Euclidean field theory the absence of a
simple action might be a problem.  In the last years we have
developed a method based on the Hamiltonian formalism (4).  Here
we will present this version of RG.  In particular we will use
this method to solve the sine-Gordon problem in $1+1$ dimension
which has a Kosterlitz-Thouless transition. \\

\begin{large}
II. Order Parameter \\
\end{large}

The magnetization, the density, or the $U(1)$ phase represent the
order parameter and characterizes the phase transition.  In all
these problems we are interested in the behavior at long distance.
As a result many different microscopical problem might have the
same long distance (low-energy) behavior.  We will assume that we
have an ultraviolet cutoff $\Lambda$, reducing $\Lambda$ to
$\Lambda' = \Lambda/s$, $s > 1$ such that the measure remains
invariant,
\begin{equation} T \exp \left[ - \frac{i}{\hbar}
\int^\infty_{-\infty} dt \int d^d x h[\phi(x,t), P(x,t)] \right]
\end{equation}
and we can construct the renormalized hamiltonian ($h[\phi,P]$ is
the hamiltonian density, "$\phi$" is the order parameter, "$P$" is
the canonical momentum $ [ \phi(x,t), $ $P(x',t) ] = i
\delta_\Lambda(x-x') $ ).  The long distance behavior is obtained
from the existence of fixed point (a set of coupling constants
which are unchanged under $\Lambda \rightarrow \Lambda/s$) \\

\begin{large}
III. Microscopic Models \\
\end{large}

We will be interested in application to strongly correlated
systems $ H = \int d^d x h[\phi(x), P(x)]$, where $h = h_0 + h_I$,
$h_0$ is the free apart and $h_I$ is the interacting part.
\begin{equation}
h_0[\phi(x), P(x)] = \frac{v}{2} [ K P^2(x) + \frac{1}{K}
(\partial_x \phi)^2]
\end{equation}
for $d > 1$, $(\partial_x \phi)^2 \equiv \sum_{i=1}^d (\partial_i
\phi)^2$, $x \equiv (x_1, \cdots x_d)$. The Ising model is
investigated choosing $K = 1$ and $h_I[\phi(x)] = \frac{1}{2} U_2
\phi^2(x) + \frac{U_4}{4!} \phi^4(x)$.  Interacting electrons in
$1+1$ dimensions are described by $d=1$, $0<K<2$ with
$h_I[\phi(x)] = U \cos(\sqrt{8 \pi} \phi(x))$ with a
Kosterlitz-Thouless transition phase characterized by $U^* = 0$
with a continuous line of fixed points $K^*$.  For $U(s)
\rightarrow 0$
we obtain a set of Luttinger liquids characterized by $0<K<2$. \\

\begin{large}
IV Constructing RG \\
\end{large}

Our starting point is the hamiltonian density, $h[\phi(x), P(x)] =
h_0[\phi(x), P(x)] + h_I[\phi(x)]$, where $h_0[\phi(x), P(x)]$ is
given in eq. (2) and $h_I[\phi(x)] = \sum_i U_i O_i(x)$ with
$O_i(x)$ being monomial of $\phi(x)$.  The fields $\phi(x) \equiv
\phi_\Lambda(x)$ and $P(x) \equiv P_\Lambda(x)$ are restricted by
$\Lambda$.\\

The RG procedure is based on the following steps:\\

(a) Project out states in the momentum interval $\Lambda/s \leq
|q| < \Lambda$ using the ''Heisenberg picture'', choose $ s = \exp
\ell$, with $\ell \rightarrow 0$ (infinitesimal).  The projection
is done such that $T \exp[ -\frac{i}{\hbar}\int dt \int d^dx
h[\phi(x,t), P(x,t)] ]$ remains invariant.

(b) In the second step the cutoff $\Lambda/s$ is restored to
$\Lambda$ by increasing the unit length by a factor of $s$,
$x'=x/s$, $t' = t/s$.

(c) In the last step we determine the scaling dimension of
$\phi(x)$ and $P(x)$ such that $\int dt \int d^d x h_0[\phi(x,t),
P(x,t)]$ remains invariant: $ \phi_{\Lambda/s}(x,t) =
\phi_{\Lambda/s}(x's, t's) = s^\Delta \phi_\Lambda(x',t')$ and $
P_{\Lambda/s}(x,t) = s^{\Delta - 1} P_{\Lambda}(x', t') $
"$\Delta$" is the scaling dimension of $\phi(x)$.  From the
invariance of the free part (eq. (2)) we obtain $\Delta =
\frac{1}{2}(d+1) - 1$. Using the scaling dimension of $\phi(x)$ we
find the scaling dimension of the monomials, $O_{i,
\Lambda/s}(x,t) = s^{\Delta_i} O_{i, \Lambda}(x', t')$.  This
leads to the scaling of the coupling constants $U_i(s) =
s^{d+1-\Delta_i} U_i(s = 1)$.  When $d+1-\Delta_i > 0$, $U_i$ is
relevant and flows to $\infty$ for $s \rightarrow \infty$. When $d
+ 1 - \Delta_i < 0$, $U_i$ is irrelevant, $U_i \rightarrow
0$. When $d+1-\Delta_i = 0$, $U_i$ is marginal.\\
\newline

In order to show how the steps (a) to (c) are implemented we will
consider a model in $1+1$ dimension ($d=1$). \\

We first separate out the fast modes from the slow ones in the
field $\phi(x)$ as well as its canonical conjugate $P(x)$:
$\phi(x) = \phi^{<}(x) + \delta \phi(x)$; $ P(x) = P^{<}(x) +
\delta P(x)$. We assumed that there is a momentum cut-off
$\Lambda$.  The slow parts, $\phi^{<}$ and $P^{<}$, contain those
components with momenta $|q| \leq \Lambda/s$, while the fast ones,
$\delta \phi$ and $\delta P$, hold the contributions from the
momentum shell, $\Lambda/s < |q| \leq \Lambda$, As a result we
find
\begin{equation}
H[\phi,P] = H[\phi^< + \delta \phi, P^< + \delta P] \equiv
H[\phi^<, P^<] + \delta H[\phi^<, \delta \phi; P^<, \delta P]
\end{equation}
where
\begin{eqnarray} \nonumber
\delta H[\phi^<, \delta \phi; P^<, \delta P] =
\int^\infty_{-\infty}dx \{ \frac{v K}{2} [\delta P(x)]^2 +
\frac{v}{2 K} [ \partial_x \delta \phi(x)]^2 \\
 + h'_I[\phi^<] \delta \phi(x) + \frac{1}{2} h''_I[\phi^<][\delta
\phi(x)]^2 + \cdots \}
\end{eqnarray}
we keep only $ ( \delta \phi(x) )^n$, $ n \leq 2 $ since we work
in the limit $ s \rightarrow 1 $.\\

To find out the effective hamiltonian of the slow modes, we need
to integrate out the fast modes defined in the momentum shell. In
the hamiltonian language we have to take the expectation value
with the "fast mode" ground state at $t = 0$, $|\Psi^>(t=0)>
\equiv | GS^> > $.  This corresponds to the ''Heisenberg picture''
with time dependent operators $\delta \phi(x,t)$,$\delta P(x,t)$.
We can achieve this by a unitary transformation such that the
linear coupling between the $\delta \phi$ field with the slow
modes $\phi^<$ is removed. (The fields $\phi^<(x)$, $P^<(x)$,
remain in the Schr\"{o}dinger picture, only $\delta \phi(x,t)$ and
$\delta P(x,t)$ are in the ''Heisenberg picture''.) This can be
carried out by introducing the following unitary transformation
$U$:
\begin{eqnarray}
\delta P'(x) = U \delta P(x) U^\dagger = \delta P(x) + \Pi(x) \\
\delta \phi'(x) = U \delta \phi(x) U^\dagger = \delta \phi(x) +
r(x)
\end{eqnarray}
$r(x)$ and $\Pi(x)$ are chosen such that the linear term disappear
from the transformed hamiltonian.  Using the Heisenberg eq. of
motion we have
\[
\partial_t \delta \phi(x,t) = v K \delta P(x,t)
\]
Demanding $[ \delta \phi(x,t), \delta P(x',t)] = [ \delta
\phi'(x,t), \delta P'(x',t)] $, we obtain that, $\Pi(x)$ and
$r(x)$ are not independent: $ \partial_t r(x,t) = v K \Pi(x,t)$.
We choose $U = \exp [ \frac{i}{2} \int_{-\infty}^{\infty} dx [
r(x) \delta P(x) - \Pi(x) \delta \phi(x) ] ] $.  By applying the
unitary transformation, we get
\begin{eqnarray} \nonumber
\int dt \delta H' = \int dt \int dx \{ \frac{v K}{2} [ \delta
P(x,t)]^2 + \frac{v}{2 K} [ \partial_x \delta \phi(x,t) ]^2 \\
\nonumber + \frac{1}{2} h''_I (\phi^<(x)) [ \delta \phi(x,t)]^2 +
\delta \phi(x,t) [ \hat{L} r(x,t) + h'_I(\phi^<(x))] \\  + r(x,t)
[ \frac{1}{2} \hat{L} r(x,t) + h'_I(\phi^<(x)) ] \}
\end{eqnarray}

In eq. (7) we define: $\hat{L} = \frac{1}{v K} \partial_t^2 -
\frac{v}{K} \partial_x^2 + h''_I(\phi^<(x))$.  To set the $\delta
\phi$ term zero, we need; $\hat{L} r(x,t) = - h'_I(\phi^<(x))$.
This gives the formal solution $ r(x,t) = - \int dy \int dt'
G(x,y; t - t') h'_I (\phi^<(y,t')) $ where $G(x,y; t-t')$ is the
Green's function of the operator $\hat{L}$, $ \hat{L}
G_\Lambda(x,t; t,t') = \delta(x-y) \delta(t-t') $.  Since
$\phi^<(x)$ is time independent $r(x)$ is determined only by the
spatial Green's function, $\hat{G}_\Lambda(x,y) \equiv
\int_{-\infty}^{t} dt' G_\Lambda(x,y; t-t')$. By substituting the
solution into eq. (7) we find
\begin{eqnarray} \nonumber
\int dt \delta H' = \int dt \int dx \{ \frac{v K}{2} ( \delta
P(x,t))^2 + \frac{v}{2 K} ( \partial_x \delta \phi(x,t))^2  +
\frac{1}{2} h''_I (\phi^<)(\delta \phi(x,t))^2\\
 - \frac{1}{2} \int dy h'_I (\phi^<(x))
\hat{G}_\Lambda(x,y) h'_I(\phi^<(y)) \}
\end{eqnarray}
Eq. (8) is exact and the Green's function $G_\Lambda(x,y;t,t')$ is
restricted to the momentum shell.  $G_\Lambda(q,\omega)$ is the
Fourier transform of $G(x,y;t,t')$ with $\Lambda/s \leq |q| \leq
\Lambda$ and $ -\infty \leq \omega \leq \infty$. Since only high
momenta in the shell are involved, we replace $\hat{G}_\Lambda$ by
$\hat{G}^{(0)}_\Lambda$ (the free Green's function). Now we
average $\delta H'$ over the ground state of the fast modes. We
use $s = \exp \ell$, $\ell \rightarrow 0$ and find:
\begin{eqnarray}\nonumber
\int dt \Delta H \equiv \int dt < GS^> | \delta H'| GS^>> = \int
dt \int dx [ \frac{K \ell}{4 \pi} h''_I(\phi^<(x)) \\ -
\frac{1}{2} \int dy h'_I (\phi^<(x)) \hat{G}_\Lambda^0(x,y)
h'_I(\phi^<(y)) ]
\end{eqnarray} \\

\begin{large}
V. Applications to the Sine-Gordon model \\
\end{large}

We choose $h_I(\phi^<) = U \cos(\sqrt{8 \pi} \phi) $ we substitute
into eq. (9) and add to eq. (9) the term $H[\phi^<, P^<]$ (see eq.
(3)).  As a result we find the effective hamiltonian
\begin{eqnarray}
H^< = \int_{-\infty}^{\infty} dx \{ \frac{v K}{2} (P^<)^2 +
\frac{v}{2 K} [ 1 + (\frac{K}{v})^2 C \frac{U^2}{\Lambda^4}\ell ]
(\partial_x \phi^<(x))^2  \\
+ U (1 - 2 K \ell) \cos(\sqrt{8 \pi} \phi^<(x)) \}
\end{eqnarray}
We see that the hamiltonian has the same form as the original one.
The superscript "$<$" indicate that the cut-off of the current
model is $\Lambda/s$ and $ C \approx 1$. To compare with the
original system, we need to restore the momentum cut-off to
$\Lambda$ by appropriately rescaling the fields and the
parameters. It is advantageous to work with the dimensionless
coupling constant $g$, $ U = g \Lambda^2$. So by rescaling the
momentum, $k \rightarrow k/s$, or in real space, $x \rightarrow x
s$, $t \rightarrow t s $ and demanding that $\int dt H_0^<$ is
invariant, we get the new set of parameters $\frac{v'}{K'} =
\frac{v}{K} [ 1 + C (\frac{v}{K})^2 \frac{u^2}{\Lambda^4}\ell ]$
in terms of the old ones: $v' K' = v K$  and using $s =
\exp(\ell)$ with $\ell$ a positive infinitesimal, we get the
$\beta$-functions:
\begin{equation}
\frac{d( v K)}{d \ell} =0; \frac{d( v / K)}{d \ell} = C g^2(v /
K)^{-1} ; \frac{d g}{d \ell} = 2 g ( 1-K)
\end{equation}
An non-trivial fixed point is observed to locate at $K=1$ and
$g=0$.  Near the point we can define a new set of variables $ x =
2 (K-1); y = \sqrt{\frac{C}{2 v_0^2}} g $, $v_0$ is the fermi
velocity before scaling $v(\ell=0)= v_0$.  The $\beta$-functions
becomes \\ $ \frac{d x}{d \ell} = - y^2$; $\frac{d y}{d \ell} = -
x y $ with the solution $x^2(\ell) - y^2(\ell) = x^2(0) - y^2(0)
\equiv const$. The scaling equation for the speed $v$ is not
necessary since the product $v(\ell) K(\ell)$ is a constant.  The
value for $v(\ell)$ can be found trivially by $ v(\ell) =
\frac{v(\ell=0)
K(\ell = 0)}{K(\ell)} $.\\

The RG flow determined by eq (12) was first reached by Kosterlitz
and Thouless (5).  There are two trivial fixed points,
corresponding to the cases $y$ goes to plus/minus infinity.  In
the fermion language the limit $y \rightarrow + \infty$
corresponds to the infinite repulsion for particles on neighboring
lattice and it is obvious at half-filling the corresponding phase
is the Mott insulator.  In the opposite limit $y \rightarrow -
\infty$, the corresponding phase at half-filling is a band
insulator in which the unit cell of the lattice is doubled and a
gap is formed at the surface of the fermi sea.  There is a third
phase which is described by the line $y = 0$ on the flow diagram.
This is the Luttinger liquid, or the spin-liquid phase.  The part
with $x > 0$ is a stable liquid phase while the $x < 0$ part is
unstable. \\

\begin{large}
ACKNOWLEDGEMENT \\
\end{large}

This work has been supported by DOE grant NO. DE-FG02-01ER45909.
\\

\begin{large}
REFERENCE
\end{large}

\begin{enumerate}
\item Fisher M.E. 1974, Rev. Mod. Phys. 4. 597; Kadanoff L.P.
1965, Physica 2, 263

\item Wilson K. G., 1975, Rev. Mod. Phys. 47, 773.

\item Anderson, P.W. and G. Yuval, 1970, Phys. Rev. Lett. 23,
1989.

\item Schmeltzer, D. 1998, Phys. Rev. B58, 69.

\item Kosterlitz J. M. and Thouless D. J. 1973, J. Phys. C6, 1181.

\end{enumerate}

\end{document}